\documentclass[10pt,letterpaper,twocolumn]{scrartcl}
\usepackage{color}

\usepackage[utf8]{inputenc}
\usepackage[english]{babel}
\usepackage{siunitx}
\usepackage{graphicx}
\usepackage[left=2cm, right=2cm]{geometry}
\usepackage{booktabs}
\usepackage{caption}
\usepackage[labelformat=empty]{subfig}
\usepackage{amsfonts, amssymb}
\usepackage[nosumlimits]{amsmath}
\usepackage{authblk}
\usepackage{url}

\usepackage[numbers]{natbib}

\begin{document}
\title{Comparing Image Quality in Phase Contrast sub$\mu$ X-Ray Tomography - A Round-Robin Study}
\author[1,2]{S. Zabler\footnote{simon.zabler@iis.fraunhofer.de}}
\author[1]{M. Ullherr\footnote{maximilian.ullherr@physik.uni-wuerzburg.de}}
\author[2]{C. Fella}
\author[2]{R. Schielein}
\author[3]{O. Focke}
\author[4]{B.~Zeller-Plumhoff}
\author[5]{P. Lhuissier}
\author[6]{W. DeBoever}
\author[1,2]{R. Hanke}
\affil[1]{Chair of X-Ray Microsopy LRM, Faculty of Physics and Astronomy, University Würzburg, Germany}
\affil[2]{Fraunhofer Development Center X-ray Technology EZRT, Josef-Martin Weg 63, 97074 Würzburg, Germany}
\affil[3]{Faserinstitut Bremen FIBRE e.V., Am Biologischen Garten 2, 28359 Bremen, Germany}
\affil[4]{Helmholtz Center for Materials and Coastal Research, Max-Planck-Straße 1, 21502 Geesthacht, Germany }
\affil[5]{INP Grenoble SIMaP - GPM2, 101 rue de la Physique, 38402 Saint Martin d'Hères, France}
\affil[6]{Bruker microCT, Kartuizersweg 3B, 2550 Kontich, Belgium}


\maketitle

\subsection*{Abstract}

How to evaluate and compare image quality from different sub-micrometer (sub$\mu$) CT scans? A simple test phantom made of polymer microbeads is used for recording projection images as well as 13 CT scans in a number of commercial and non-commercial scanners.
From the resulting CT images, signal and noise power spectra are modeled for estimating volume signal-to-noise ratios (3D SNR spectra).
Using the same CT images, a time- and shape-independent transfer function (MTF) is computed for each scan, including phase contrast effects and image blur ($\mathrm{MTF_{blur}}$).
The SNR spectra and MTF of the CT scans are compared to 2D SNR spectra of the projection images. In contrary to 2D SNR, volume SNR can be normalized with respect to the object's power spectrum, yielding detection effectiveness (DE) a new measure which reveals how technical differences as well as operator-choices strongly influence scan quality for a given measurement time. Using DE, both source-based and detector-based sub$\mu$ CT scanners can be studied and their scan quality can be compared. Future application of this work requires a particular scan acquisition scheme which will allow for measuring 3D signal-to-noise ratios, making the model fit for 3D noise power spectra obsolete.



\section{Introduction}

How should we judge and compare image quality in sub$\mu$ computed tomography (CT) scans?
The term "sub$\mu$" designs CT scans which a. employ spatial (voxel) sampling in the sub-micrometer range, i.e. $100\,$nm to $999\,$nm, and b. achieve spatial resolutions which justify this sampling.
While the number of installations of sub$\mu$ scanners keeps on growing (many in materials science laboratories) discussions about scan quality rarely surpass the visual comparison of fancy pictures. Medical imaging traditionally offers many quantitative measures which could be applied to turn the comparison of different sub$\mu$ CT scanners into a more scientific argument. Modulation transfer functions (MTF), signal-to-noise ratio (SNR, commonly defined as the temporal mean divided by the standard deviation) or detective quantum efficiency (DQE) all have their justification and usefulness for evaluating the performance of an imaging system such as CT, but each measure only covers a specific part and not the device in total \cite{Ullherr:2019,Cunningham:1998}. Furthermore sub$\mu$ CT scanners employ contrast mechanisms which are fundamentally different from medical CT scanners.

\subsection{The origin of polychromatic phase contrast}

The most prominent difference between medical CT and sub$\mu$ CT is X-ray inline phase contrast. The latter stems from Fresnel-type near-field diffraction and was first reported by the inventors of the X-ray shadow microscope, Cosslett and Nixon \cite{Goodman:2005, Cosslett:1951}. Phase contrast images show Fresnel-fringes which highlight microscopic material interfaces and enhance structural details.
The fringe-visibility requires spatial coherence and a sufficiently long propagation length $z$ \cite{Pelliccia:2011}.
These criteria demand for high spatial resolution which has to be of the order of the fringe-spacing $\sqrt{\lambda z}$ ($\lambda$: X-ray wavelength). This criterion can be either met by small focal spots (source-based scanners) or by high-resolution detectors (detector-based scanners). Both employ cone-beam geometry for which the propagation length $z$ is given by \cite{Guigay:1977}:
\begin{equation}
z=\frac{z_\mathrm{SOD}\cdot z_\mathrm{ODD}}{z_\mathrm{SOD}+z_\mathrm{ODD}}= \frac{z_\mathrm{ODD}}{M}
\end{equation}
with $z_\mathrm{SOD}$ the source-object distance (SOD), $z_\mathrm{ODD}$ the object-detector distance (ODD) and $M$ the geometric magnification. We define two limit cases. First, source-based scanners:
\begin{equation}
z_\mathrm{SOD}\ll z_\mathrm{ODD} \Rightarrow z\approx z_\mathrm{SOD}
\end{equation}
Second, detector-based scanners:
\begin{equation}
z_\mathrm{SOD}\gg z_\mathrm{ODD} \Rightarrow z\approx z_\mathrm{ODD}
\end{equation}
For the symmetric case ($M=2$) $z$ is maximal:
\begin{equation}
z_\mathrm{SOD}=z_\mathrm{ODD}\Rightarrow z=z_\mathrm{SOD}/2
\end{equation}

Sub$\mu$ CT scanners are different from medical CT and industrial inspection (NDT scanners). The latter typically spatially resolve hundred micrometers at best and they employ much higher photon energies (implying weaker phase contrast) for penetrating large objects. Under these conditions phase contrast is generally not observed.
The latter can however be realized for moderate resolutions by increasing $z$. This has been proposed recently for medical applications using $20\,\mu$m spatial sampling and $z=408\,$mm at $\lambda_\mathrm{avg}\approx 0.02\,$nm average wavelength \cite{Ghani:2019}.

Wilkins et al., who took up Nixon's and Cosslett's concept of the X-ray shadow microscope, stated correctly that monochromaticity (temporal coherence) is no stringent requirement for observing X-ray phase contrast \cite{Wilkins:1996}. Polychromatic phase contrast is an inherent feature of all sub$\mu$ CT scanners provided they have sufficient spatial resolution (few micrometers or better) and propagation length (few millimeters or more) \cite{Zabler:2012}.

\subsection{Are all sub$\mu$-CT scanners similar?}

Two different concepts of scanners have evolved. While Carl Zeiss (Versa scanners, former X-radia Inc.) and Rigaku (former Reflex Ltd.) advertise detector-based sub$\mu$ CT \cite{Tous:2008, Feser:2008}, General Electric (nanotom scanners, former Phoenix X-ray GmbH), Bruker (former Skyscan BE) and RX Solutions SA (Easytom scanners) produce source-based systems.

Traditionally source-based systems feature a small focal spot (typically $<2\,\mu$m) and strong geometric magnification, similar to Cosslett's Shadow Microscope. However, many micro-focal systems employ CCD detectors (RXS and Bruker) which imply a lesser magnification. Detector-based scanners feature larger focal spots, higher anode power, and geometric magnification $M<2$ \cite{Fella:2017}.
These systems employ high-resolution microscope detectors which were pioneered by Andreas Koch \cite{Koch:1998}.
However, in order to enlarge their field-of-view these detectors may employ optics and screens with lower resolution, leading to $M>2$ (e.g. Zeiss Versa 4x). Some include additional Flatpanel (FP) detectors.

Both scanner types have their benefits and drawbacks. Producing a small focal spot requires micrometer-thin transmission targets. The heat-load as well as the electron transparency are the two factors limiting the brightness of micro-focal X-ray sources. Microscope X-ray detectors employ microscopic-thin scintillator screens (single crystal or polycristalline) for converting X-rays to visible light. These screens preferably convert softer X-rays ($<20\,$keV) while harder X-rays mostly pass through. Light collection efficiency is limited by the aperture of the microscope lens.


\subsection{Evaluating CT image quality}

Optimizing either MTF (modulation transfer function) or temporal SNR generally does not yield the best image quality. For a detailed review of the underlying problem refer to \cite{Ullherr:2019}.
Using SNR spectra ($\mathrm{SNR}(u)$; note that $u$ refers to the radial spatial frequency) as a quality measure comes naturally. The latter include modulation transfer, phase contrast and noise effects all with respect to an object at hand. They can be measured easily from a series of transmission images, and we demonstrated their use by finding the optimum magnification for phase contrast sub$\mu$ imaging in the LMJ-CT setup \cite{Ullherr:2018}.

While the universality of $\mathrm{SNR}(u)$ can be questioned because it depends on the object's power spectrum, using the same or perfectly similar objects allows for comparing the performance of different imaging systems (or different parameter settings in one system) by measuring and comparing their polychromatic SNR spectra.

Admittedly, evaluating CT scanners should include CT images. Therefore extending $\mathrm{MTF}(u)$ and $\mathrm{SNR}(u)$ to 3D is desirable.
Three-dimensional object coordinates are meant by "3D" (real- and reciprocal space), whereas "2D" refers to horizontal and vertical axes in transmission images.
Furthermore, we introduce detection effectiveness (DE) which includes MTF and noisiness while being independent of the object's structure. $\mathrm{DE}(u)$ requires exact knowledge of the object's power spectrum $S_\mathrm{object}(u)$. This study uses a simple and reproducible test object (mixture of two sizes of microbeads) in order to calculate $\mathrm{MTF_{3D}}$, $\mathrm{SNR_{3D}}$ and $\mathrm{DE_{3D}}$ for each scanner and for each parameter setting.
We thereby approximate CT imaging to be a linear time-invariant (LTI) system with a single real Fourier kernel that applies to any object of similar X-ray interaction. This approximation implies isotropy while neglecting non-linearities and local effects which arise during back-projecting beam intensities, known as fan-beam artifacts \cite{Buzug:2011}. We argue that whenever CT is applied for materials characterization, isotropy is assumed while local effects are ignored by image analysis.

Except for the hint that no attenuation filter was necessary we did not give precise instructions to the machine operators. Instead we wanted them to parametrize and perform the scans in the way they considered optimal. This test therefore invokes not only different scanners but also significantly different parameter settings, sometimes on the same scanner. Admittedly, "optimal" leaves at least two choices open: a. employing the scanners' best resolving power (best MTF), or b. making the best CT image of the phantom in the shortest scan time (best SNR).
In the outcome of this study we shall see that both choices were made by different operators. Consequently, those scans with an optimal $\mathrm{MTF_{3D}}$ are distinct from those with better $\mathrm{SNR_{3D}}$ and $\mathrm{DE_{3D}}$.

\section{Materials and Methods}

Table \ref{tab:scanners} lists the different CT scanners and settings which were used to scan the test object and their measurement settings.

In addition to the commercial scanners, this study includes three systems which have been developed by the Fraunhofer Development Center X-ray Technology (EZRT): The ntCT (based on Excillum's nanotube source), the LMJ-CT (based on Excillum's Liquid-Metal-Jet source) and the new compact Click-CT scanner \cite{Fella:2017}.
\begin{table*}[!th]
\centering
\vspace{2mm}
\begin{tabular}{l|l|l}
\hline
Scanner / Operator & Source / Scan geometry & Detector and scan parameters\\
\hline
\multicolumn{3}{c}{Detector-based systems}\\
\hline
\textbf{Versa 520 20x FIBRE}
& Transmission $80\,$kV, $88\,\mu$A
& Microscope $0.675\,\mu$m/pixel \\
Oliver Focke & $\mathrm{SOD}=22.15\,$mm, $\mathrm{SDD}=30\,$mm
& $\Delta x=0.498\,\mu$m, $t=3201\ast 12\,$s \\
\hline
\textbf{Versa 520 4x FIBRE}
& $80\,$kV, $87.5\,\mu$A
& Microscope $3.375\,\mu$m/pixel \\
Oliver Focke
& $\mathrm{SOD}=12.3\,$mm, $\mathrm{SDD}=80\,$mm
& $\Delta x=0.518\,\mu$m $t=3201\ast 3\,$s\\
\hline
\textbf{Versa 520 20x KIT}
& Transmission $80\,$kV, $88\,\mu$A
& Microscope $0.675\,\mu$m/pixel \\
Jochen Joos 
& $\mathrm{SOD}=85.10\,$mm, $\mathrm{SDD}=96.5\,$mm
& $\Delta x=0.595\,\mu$m, $t=3201\ast 10\,$s\\
\hline
\textbf{Versa 520 4x KIT}
& $80\,$kV, $87.5\,\mu$A
& Microscope $3.375\,\mu$m/pixel \\
Jochen Joos
& $\mathrm{SOD}=15.5\,$mm, $\mathrm{SDD}=79.9\,$mm
& $\Delta x=0.657\,\mu$m, $t=1601\ast 4\,$s\\
\hline
\textbf{LMJ-CT FhG}
& Excillum Liquid-Metal-Jet $70\,$kV, $1.43\,$mA
& Microscope $0.62\,\mu$m/pixel\\
Simon Zabler
& $\mathrm{SOD}=165\,$mm, $\mathrm{SDD}=185\,$mm
& $\Delta x=0.553\,\mu$m, $t=1400\ast 8\,$s\\
\hline
\textbf{Click-CT 20x FhG}
& Hamamatsu L12161 $40\,$kV, $250\,\mu$A
& Opt.Peter XRM $0.325\,\mu$m/pixel \\
Simon Zabler
& $\mathrm{SOD}=38.5\,$mm, $\mathrm{SDD}=43.3\,$mm
& $\Delta x=0.289\,\mu$m, $t=2400\ast 10\,$s\\
\hline
\multicolumn{3}{c}{Source-based systems}\\
\hline
\textbf{nanotom m U WEI}
& Transmission $90\,$kV, $150\,\mu$A
& Flatpanel  $100\,\mu$m/pixel\\
Franziska Vogt
& $\mathrm{SOD}=3.83\,$mm, $\mathrm{SDD}=600\,$mm
& $\Delta x=0.637\,\mu$m, $t=5001\ast 3.75\,$s\\
\hline
\textbf{nanotom s HZG}
& Transmission $80\,$kV, $90\,\mu$A
& Flatpanel $50\,\mu$m/pixel \\
Berit Zeller-Plumhoff
& $\mathrm{SOD}=3.64\,$mm, $\mathrm{SDD}=210\,$mm 
& $\Delta x=0.867\,\mu$m, $t=3001\ast 16\,$s\\
\hline
\textbf{Easytom CCD INPG}
& Hamamatsu L10711 LaB6, $40\,$kV, $120\,\mu$A
& P.I. Quad-RO CsI $24\,\mu$m/pixel\\
Pierre Lhuissier
& $\mathrm{SOD}=1.85\,$mm, $\mathrm{SDD}=72.9\,$mm
& $\Delta x=0.609\,\mu$m, $t=1600\ast 5\,$s\\
\hline
\textbf{Easytom CCD RXS}
& Hamamatsu L10711 LaB6, $100\,$kV, $162\,\mu$A
& P.I. Nano-XF CsI $2\times 9\,\mu$m/pixel\\
Solene Valton
& $\mathrm{SOD}=3.17\,$mm, $\mathrm{SDD}=142\,$mm
& $\Delta x=0.402\,\mu$m, $t=1568\ast 30\,$s\\
\hline
\textbf{Easytom FP RXS}
& Hamamatsu L10711 LaB6, $100\,$kV, $171\,\mu$A
& Varian 2520DX CsI $127\,\mu$m/pixel\\
Solene Valton
& $\mathrm{SOD}=3.07\,$mm, $\mathrm{SDD}=971\,$mm
& $\Delta x=0.402\,\mu$m, $t=1440\ast 30\,$s\\
\hline
\textbf{Skyscan 2214 Bruker}
& Hamamatsu L10711 W, $60\,$kV, $200\,\mu$A
& Ximea CCD GdOS $2\times 9\,\mu$m/pixel \\
Wesley DeBoever
& $\mathrm{SOD}=7.2\,$mm, $\mathrm{SDD}=244.5\,$mm
& $\Delta x=0.53\,\mu$m, $t=1801\ast 8.75\,$s\\
\hline
\textbf{ntCT FhG}
& Excillum nanotube $60\,$kV, $100\,\mu$A
& Dectris Säntis CdTe $75\,\mu$m/pixel\\
Antonia Ohlmann
& $\mathrm{SOD}=1.45\,$mm, $\mathrm{SDD}=423.5\,$mm
& $\Delta x=0.256\,\mu$m, $t=2400\ast 25\,$s\\
\hline
\end{tabular}
\caption{Scanners, operators, source and detector parameters for the sub$\mu$ CT scans. $\Delta x$ refers to voxel sampling, $t$ to total exposure time. FIBRE is part of MAPEX Center for Materials and Processes at University Bremen; KIT: Karlsruhe Institute of Technology; FhG: Fraunhofer-Gesellschaft; U WEI: Bauhaus-Universität Weimar; HZG: Helmholtz-Zentrum Geesthacht Center for Materials and Coastal Research; INPG: Institut National Polytechnique de Grenoble; RXS: Rayons X Solutions SA; Bruker MicroCT Kontich, Belgium.}\label{tab:scanners}
\end{table*}

\subsection{2D SNR from transmission images}

For most scanners we were able to measure $\mathrm{SNR_{2D}}(u)$ from series of transmission images before proceeding with the CT scan. Table \ref{tab:SNR-measures} lists the scanners and their settings (same as for the CT scans) for these measurements. For the X-radia Versa 520 scanner at KIT (4x and 20x lens) we computed $\mathrm{MTF_{2D,blur}}(u)$ from images of a slanted edge (see Appendix).
This procedure was however only applied to one scanner while noise power spectra (NPS) $N_\mathrm{2D}(u)$ were used to estimate the (noise effective) blur and the total X-ray conversion of the remaining systems.

According to \cite{Ullherr:2018} $\mathrm{SNR}_{\mathrm{2D},\tau}$ is computed from a series of $K$ transmission images of the test object with exposure time $\tau$.
First, the detected intensity images $I_{\tau,j}(x,y)$ are normalized with respect to averaged \emph{flatfield} images $I_\mathrm{flat,avg}(x,y)$ and detector dark current $I_\mathrm{dark,avg}(x,y)$, $j\in 1\dots K$:
\begin{equation}
d_{\tau,j}=-\ln\left[\frac{I_{\tau,j}-I_\mathrm{dark,avg}}{I_\mathrm{flat,avg}-I_\mathrm{dark,avg}}\right]\label{eq:normalize}
\end{equation}
From $d_{\tau,j}$ we compute $D_\mathrm{avg}(u)$, the power spectrum of the average detected signal, and $\left\langle D_j(u)\right\rangle$ the average of the individual detected radial power spectra $D_j(u)$. Note that reciprocal coordinates $\vec{u}=(u_x,u_y,u_z)^T$ are reduced to a scalar $u=|\vec{u}|$ by radial or spherical averaging. Thus, volume power spectra can be show as line plots.

Assuming additive noise , i.e. $d_{\tau,j}=s_\tau+n_{\tau,j}$, we can compute power spectra for deterministic signal ($s_\tau$) and random noise ($n_{\tau,j}$) separately (cf. \cite{Ullherr:2018}):
\begin{eqnarray}
S_{\mathrm{2D},\tau}(u)= \frac{D_\mathrm{avg}(u)-K^{-1}\left\langle D_j(u)\right\rangle}{1-K^{-1}}\\
N_{\mathrm{2D},\tau}(u)= \frac{\left\langle D_j(u)\right\rangle-D_\mathrm{avg}(u)}{1-K^{-1}}
\end{eqnarray}
The denominator of these two cancels out when SNR is computed (same for 2D and 3D):
\begin{equation}
\mathrm{SNR}(u)=\frac{S(u)}{N(u)}\label{eq:snrdef}
\end{equation}
The resulting 2D SNR is given by:
\begin{equation}
\mathrm{SNR}_{\mathrm{2D},\tau}=
\frac{D_\mathrm{avg}- K^{-1}\left\langle D_j\right\rangle-(1-K^{-1})A}{\left\langle D_j\right\rangle-D_\mathrm{avg}}\label{eq:SNR-measure}
\end{equation}
The correction term $A(u)$ corresponds to artificial signal contributions (e.g., arising from the detector dark current and flatfield images power spectra). It is derived in the appendix.

\subsection{3D SNR and DE of volume images}

Similar to $\mathrm{SNR_{2D}}(u)$ which is deduced from series of transmission images, measuring $\mathrm{SNR_{3D}}(u)$ would require series of 5 to 10 identical CT scans, or 5 to 10 projection images at each angular position, from which corresponding volume images and volume signal and noise power spectra can be computed. Because this acquisition scheme could not be applied due to a lack of flexibility in commercial CT scanners' software, we chose to estimate $\mathrm{SNR_{3D}}(u)$ from a single CT scan, by the following procedure:

The detected power spectrum $D_\mathrm{3D}$ of the CT contains effects of modulation transfer (MTF, i.e. blur and phase contrast) and of additive noise:
\begin{equation}
D_\mathrm{3D}=S_\mathrm{3D}+N_\mathrm{3D}=b^2\mathrm{MTF^2_{3D}}\cdot S_\mathrm{object} +N_\mathrm{3D}
\end{equation}
$b$ represents the attenuation strength, $S_\mathrm{object}$ the object's power spectrum which is determined for each scan (see section~\ref{sec:Volume}).
For the computation of $\mathrm{SNR_{3D}}$, $\mathrm{S_{3D}}$ and $\mathrm{N_{3D}}$ need to be determined. This is done by a model fit, which
for practical reasons\footnote{$S_\mathrm{object}$ varies over many orders of magnitude, thereby making a model fit of $D_\mathrm{3D}$ difficult.} is applied to the following function:
\begin{equation}
\frac{D_\mathrm{3D}}{S_\mathrm{object}}=b^2\mathrm{MTF^2_{3D}}+\frac{N_\mathrm{3D}}{S_\mathrm{object}}\label{eq:datafit}
\end{equation}
Fit parameters are determined for the models of $\mathrm{MTF_{3D}}$ and $N_\mathrm{3D}$ (cf. eq. \ref{eq:MTF-model} and \ref{eq:NPS}).
Note that for the (high) frequency band where $D_\mathrm{3D}$ is pure noise, the ratio $D_\mathrm{3D}/S_\mathrm{object}\approx N_\mathrm{3D}/S_\mathrm{object}$ resembles a ragged curve which may even increase towards high frequencies.
The model fit for $N_\mathrm{3D}$ is then used for computing $\mathrm{SNR_{3D}}$:
\begin{equation}
\mathrm{SNR_{3D}}(u)=\frac{D_\mathrm{3D}(u)-N_\mathrm{3D}(u)}{N_\mathrm{3D}(u)}\label{eq:3DSNR}
\end{equation}
The approximation which is hereby made is the model fit of the volume NPS. The latter therefore constitutes the main uncertainty for the estimation of $\mathrm{SNR_{3D}}(u)$.

In order to make $\mathrm{SNR_{3D}}$ independent of the object structure, it is normalized with respect to $S_\mathrm{object}$, yielding detection effectiveness $\mathrm{DE_{3D}}(u)$ \cite{Ullherr:2019}:
\begin{equation}
\mathrm{DE_{3D}}(u)=\frac{\mathrm{SNR_{3D}}(u)}{S_\mathrm{object}(u)}
\end{equation}
While $\mathrm{SNR_{3D}}$ is strongly governed by the structure (shape) of the object, $\mathrm{MTF_{3D}}$ and $\mathrm{DE_{3D}}$ are not.
The latter are independent of the object's structure while they remain dependent of everything else: E.g. the object's interaction strength (attenuation and phase shift) which in turn is weighted with the X-ray energy spectrum (including beam hardening). Image quality is intrinsically application-specific, and so is DE. 

\subsection{Modeling modulation transfer in CT images}

We define the modulation transfer to include both the effects of image blur and phase contrast.
Both for 2D and 3D, MTF can be written as:
\begin{equation}
\mathrm{MTF}(u) = \mathrm{MTF_{blur}}(u)\cdot \mathrm{MTF_{phase}}(u)\label{eq:MTF-model}
\end{equation}
$\mathrm{MTF_{3D,blur}}$ in eq. \ref{eq:datafit} is modeled by the product of an exponential with a Gaussian function, which is usually a good approximation.
\begin{equation}
\mathrm{MTF_{3D,blur}} = \mathcal{F}_\mathrm{3D}\left[\frac{1}{1+(r/\mu)^2} \ast \exp\left(-\frac{r^2}{2\sigma^2}\right) \right]\label{MTF-model}
\end{equation}
\begin{equation}
\varpropto \exp\left(-2\pi\mu |u|\right)\cdot \exp\left( -2\pi^2\sigma^2u^2\right)\label{eq:exponential}
\end{equation}
$r$ is the radial real space coordinate, $\mu$ and $\sigma$ are the widths of the Lorentzian and Gaussian functions respectively (eq. \ref{MTF-model}). Due to the radial symmetry of these functions eq. \ref{eq:exponential} is equally valid for 2D and 3D Fourier transforms.  Meanwhile, phase propagation is approximated by a parabola \cite{Paganin:2002}: 
\begin{equation}
\mathrm{MTF_{3D,phase}} = 1+p^2u^2\label{eq:Pagnin-model}
\end{equation}

For a given wavelength $\lambda$ the phase strength $p$ depends on the material-specific X-ray interactions as well as on the propagation length $z$ \cite{Paganin:2002,Weitkamp:2011}:
\begin{equation}
p(\lambda,z)=\sqrt{\frac{z\delta}{\mu_\mathrm{abs}}}
\end{equation}
where $\delta$ and $\mu_\mathrm{abs}$ are the interaction strengths for phase shift and attenuation, respectively. Note that eq. \ref{eq:MTF-model} results in a peaked shape of $\mathrm{MTF_{3D}}$.
While the definition of $\mathrm{MTF_{blur}}$ for 2D transmission images is the same as for 3D CT images, we shall refer to $\mathrm{MTF_{3D}}$ and $\mathrm{MTF_{3D,blur}}$ for those curves which are computed from 3D power spectra. In the appendix $\mathrm{MTF_{2D,blur}}$ is estimated from slanted edge line profiles. While the latter appear consistent with $\mathrm{MTF_{3D,blur}}$ it would be wrong to assume that both are identical (one relates to the test object, the other to a tungsten edge).

\begin{figure}[!th]
\centering
\includegraphics[width=8.0cm]{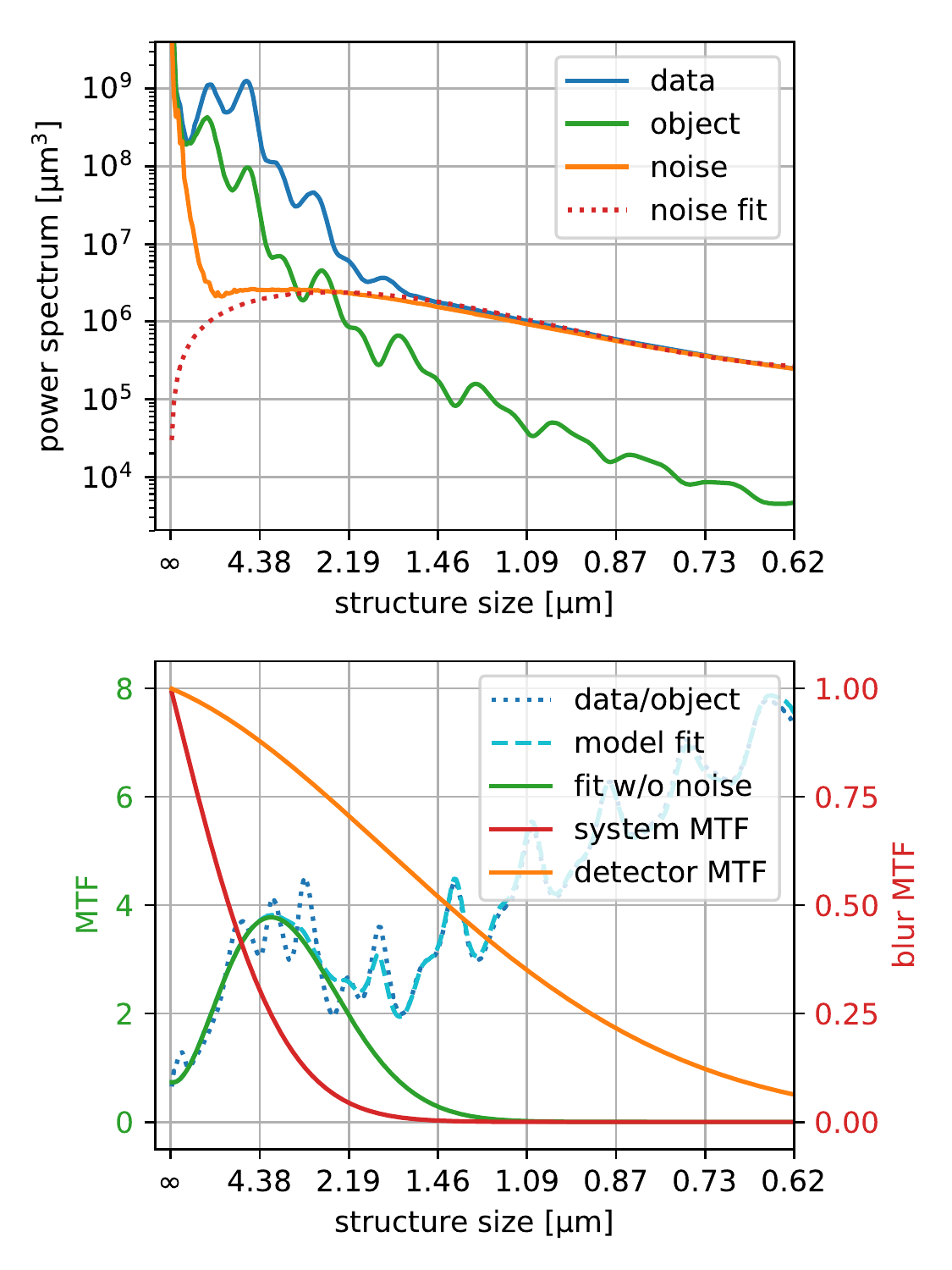} 
\caption{
Model fit for the nantom s HZG. Top: Detected signal and object spectra, volume image noise (measured from an empty volume region) and noise model fit ($c=24.2$, cf. table \ref{tab:SNR-measures}, eq. \ref{eq:NPS}). The latter was independently obtained from the model fit below. Bottom: Model fit for $\mathrm{D_{3D}}/S_\mathrm{object}$ (eq. \ref{eq:datafit}); resulting $\mathrm{MTF_{3D}}$ and system blur ($\mathrm{MTF_{3D,blur}}$). The latter is compared to (detector) noise MTF.}\label{fig:HZG-model-fit}
\end{figure}
Fig. \ref{fig:HZG-model-fit} illustrates the model fit (eq. \ref{eq:datafit}) for the example of the nanotom s (HZG) scan. The transfer function has a peaked shape according to eq. \ref{eq:MTF-model}. The system $\mathrm{MTF_{3D,blur}}(u)$ appears far worse than the (detector) noise MTF. Hence, resolution in this scan is clearly not limited by the detector but by the focal spot width which does not affect $\mathrm{MTF_{noise}}(u)$. For this particular example the object was not centered resulting in a large empty volume region which was used for measuring volume noise $N_\mathrm{3D}$ (Fig. \ref{fig:HZG-model-fit} top). This measurement inevitably contains low-frequency deterministic artifacts, yet the medium and high frequency parts match the model fit reasonably well.

Because both $\mathrm{MTF_{phase}}$  and $\mathrm{MTF_{blur}}$ are determined by the fit, the latter can be compared to $\mathrm{MTF_{noise}}$ (see section \ref{sec:model-noise}). Concerning the latter, an accurate estimate of $\mathrm{MTF_{noise}}$ requires a sufficiently large (high) frequency band in which pure noise can be estimated. Consequently, $\mathrm{MTF_{noise}}$ is less accurate for those scans with very high spatial resolution (Easytom and ntCT).

\subsection{Modeling noise in transmission and CT images\label{sec:model-noise}}

The model for the shape of the projections' noise power spectra includes a scaling factor $a$, conversion to and collection of optical photons $c$ as well as noise effective blur $\mathrm{MTF_{noise}}(u)$ \cite{Cunningham:1999}:
\begin{equation}
N_\mathrm{2D}(u)=a\cdot (1+c\cdot \mathrm{MTF^2_{noise}}(u))\label{eq:NPS2D}
\end{equation}
Here, $a$ scales with the detected intensity.
This equation can be used for estimating X-ray conversion $c$ of the detector from the DC and the Nyquist amplitudes in $N_\mathrm{2D}(u)$, i.e. by assuming $\mathrm{MTF_{noise}}(u_\mathrm{ny})\approx 0$ and $\mathrm{MTF_{noise}}(0)=1$:
\begin{equation}
c\approx\frac{N_\mathrm{2D}(0)-N_\mathrm{2D}(u_\mathrm{ny})}{N_\mathrm{2D}(u_\mathrm{ny})}
\label{eq:conversion}
\end{equation}

Note that 
$\mathrm{MTF_{noise}}$ and $\mathrm{MTF_{blur}}$ remain fundamentally different quantities. The former is a property of the detector alone whereas the latter refers to the blurring of an actual object both by source and detector. Note that $\mathrm{MTF_{noise}}\geq\mathrm{MTF_{blur}}$.

Concerning $N_\mathrm{3D}$, tomographic back-projection adds convolution and linear interpolation to the image process chain resulting in an altered model for volume NPS \cite{Hanson:1979}.
\begin{equation}
N_\mathrm{3D}(u) = |u|\cdot a' \cdot \mathrm{sinc}^{3.4}(u) \left(1+c\cdot\mathrm{MTF^2_{noise}}(u)\right)
\label{eq:NPS}
\end{equation}
This model for $N_\mathrm{3D}(u)$ was deduced from a CT reconstruction of simulated noise images. The exponent 3.4 thereby results from spherical averaging whereas an exponent 4 would result from bilinear interpolation. Note that additional corrections $O(u^6, u^{12})$ apply for the high frequencies while additional low frequency noise is caused by tomographic imaging artifacts, such as ring artifacts, etc.

\subsection{An objective test object?}

Defining an appropriate test object for evaluating different sub$\mu$ CT scanners, that is easy to replicate and still produces meaningful universal results, requires some thought. Generally, during setup and commissioning of new scanners different test structures are scanned in order to validate the image quality visually (e.g. foams, granules and organic composites). While these structures are rich in detail at different length scales and produce nice images they do not allow for determining their exact shape (ground truth).
Data power spectra can be calculated from such scans and a good noise model can even yield $\mathrm{SNR_{3D}}$ spectra. However, problems arise when object-independent descriptors are required, i.e. $\mathrm{MTF_{3D}}$ and $\mathrm{DE_{3D}}$. Consequently, scan performance can only be compared when the \emph{exact same} test object is used.

In order to allow for quick and reliable evaluation of the object structure
the latter has to be \emph{simple enough}, while maintaining a high level of detail and general isotropy.
We chose a mixture of polymer microbeads of two (monodisperse) sizes ($10\,\mu$m and $20\,\mu$m in diameter, courtesy of Ralf Nordal, Microbeads Norway). The microbeads are perfectly spherical and made of cross-linked PMMA. They were mixed $50\mathrm{wt}\%$-$50\mathrm{wt}\%$ which should yield a number ratio of $8:1$.
The mixture was poured into a thin glass capillary of $\approx1\,$mm outer diameter. Due to the common use of these capillaries in X-ray powder diffraction the wall thickness ($20\,\mu$m) and hence the X-ray transmission of the container is highly constant. The latter in combination with the inner diameter of the capillary are part of the sample definition, i.e. its polychromatic X-ray transparency.

Using this test object has several advantages:
\begin{itemize}
\item The spheres are easily detected and their size can be determined, e.g. by computing the Euclidean Distance Transform (EDT) from a binary volume image.
\item From the histogram of the sphere diameters we obtain an absolute scale for the volume image and can correct erroneous estimates of its voxel size.
\item The object is composed of a single material and therefore easy to segment after applying phase retrieval and Wiener-deconvolution to the CT scans \cite{Ullherr:2015}.
\item Computation of the virtual object structure by placing spheres at the center coordinates which are obtained from the EDT maxima is fast and accurate.
\item The noise power spectrum is monotonously decreasing, whereas the signal power spectrum shows distinctive structural peaks from the harmonics of the two sphere diameters. From the shape of the data power spectrum we can therefore easily derive scaling and distinguish signal and noise.
\item Mixing of two sizes yields good isotropy by preventing long-range ordering of the spheres which has to be expected for a single size.
\item Instead of having one unique object and carrying it around we could send out many similar objects to the users of sub$\mu$ CT scanners and we would replace them if one accidentally broke. This happened frequently.
\end{itemize}
While profiting from all these benefits we have to point out that the PMMA microbeads are quasi "pure phase" objects, hence the sample shows negligible X-ray attenuation in sub$\mu$ scan mode. The performance for higher-Z materials (e.g. metal) is different.

\subsection{Volume image processing\label{sec:Volume}}

\begin{figure}[t]
\centering
\includegraphics[width=8cm]{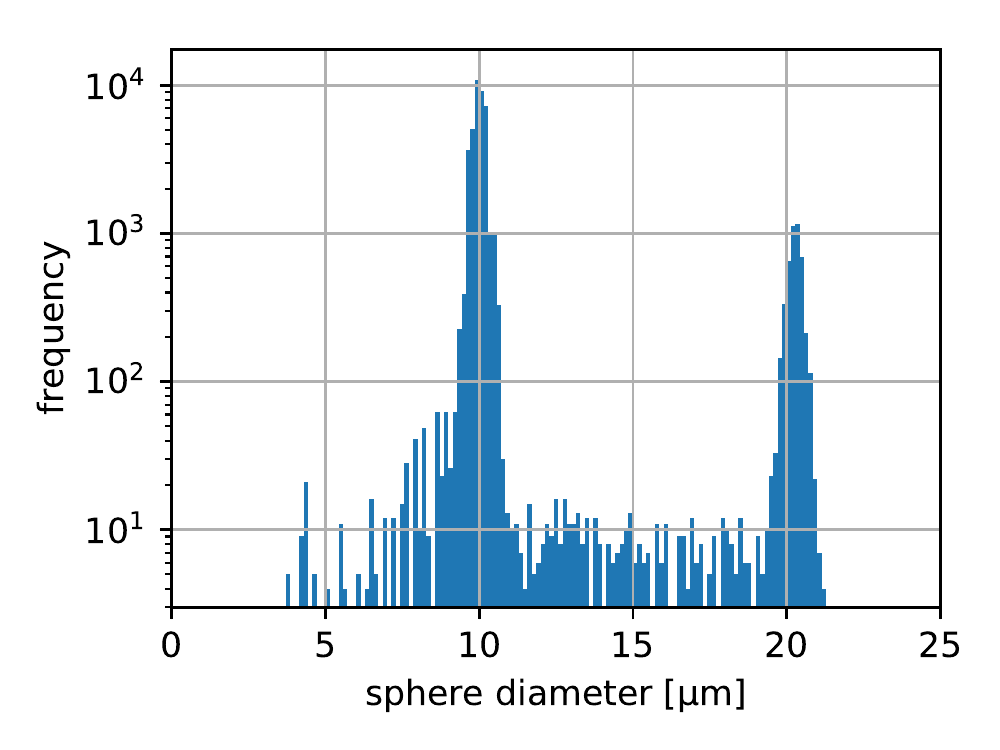}
\caption{Sphere diameter histogram for the "Versa 20x FIBRE" dataset. $\pm 20\%$ deviation were tolerated for each size population ($10\,\mu$m and $20.3\,\mu$m).}\label{fig:sizehisto}
\end{figure}

Except for three scans (RX Easytom) all volume reconstructions were done with our free \emph{python} X-ray imaging toolkit (pyXIT) \cite{pyXIT}.
The Easytom scans were reconstructed by the company's software in order to compensate for projection-wise random sample shifts. The latter are applied to compensate for detector pixel-defects which would otherwise cause ring-artifacts.
Alternatively, the latter can be corrected by standard angular median method. Presumably, all systems applied a focal spot drift correction prior to writing out the raw images. The two Versa scanners exported a reduced image size (e.g. 1994x1994 instead of 2048x2048) indicating that image distortions were corrected before the file export. None of these steps significantly affected reconstruction or image analysis.
pyXIT was used for image segmentation, Paganin-type phase retrieval and Wiener-deconvolution \cite{Paganin:2002, Ullherr:2018}. Following volume reconstruction, all those steps were applied simultaneously to each data set.
\begin{figure}[t]
\centering
\includegraphics[width=7.0cm]{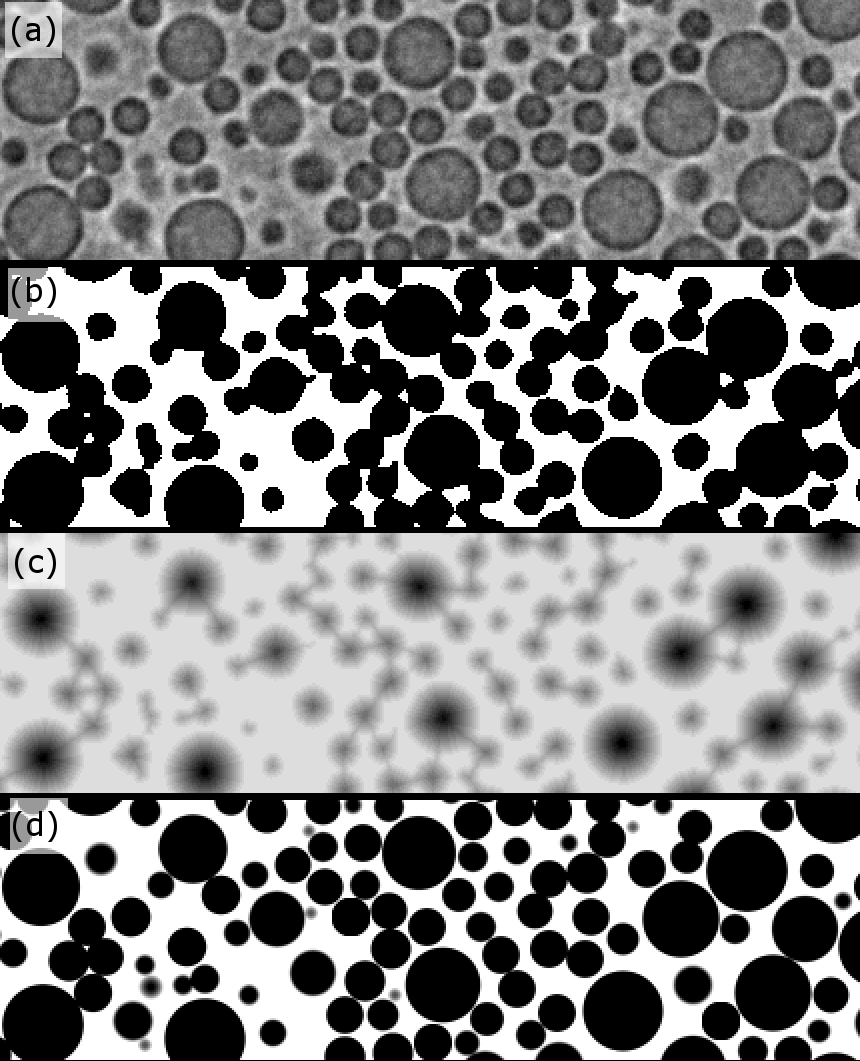}
\caption{Image processing chain to create the virtual object structure: The raw data (a) is converted to a binary volume image (b) through phase retrieval, Wiener deconvolution and threshold segmentation. The algorithm searches the EDT (c) for the spheres' center coordinates, from which (d) the object structure $O(\vec{x})$ is created.}\label{fig:imageprocess}
\end{figure}
Phase strength and the deconvolution kernel were adjusted in order to remove Fresnel-fringes and reduce noise. The volume images were then segmented using manual thresholding. We used functions from the python module \textbf{scipy.ndimage} on sub-volumes of approx. $800\times 800\times 800$ voxels size to fill holes in the foreground, then computed the EDT from the segmented images \cite{scipy}. The process is depicted in Fig. \ref{fig:imageprocess}.
A search algorithm then finds the center coordinates of all spheres (pixel precision) and lists their diameters. The histogram of the latter (Fig. \ref{fig:sizehisto}) shows two distinct peaks which correspond on average to the $10\,\mu$m and $20.3\,\mu$m spheres.
Since threshold segmentation of the somewhat noisy data causes underestimation of the spheres' diameters we added a constant (typically $\approx 1$ pixel) to the latter and scaled the voxel size accordingly to match the precise diameters.

A virtual object $O(\vec{x})$ is created by filling an empty volume with spheres at the center coordinates and with the diameters found by the sphere-search.
To include partial-voxel fill-effects the virtual spheres were over-sampled by a factor of 6 then down-sampled to the voxel size of the corresponding data set before placing them into the volume. Due to imprecision in position and size, some spheres overlapped by one pixel. We ignored this effect because the concerned volume fraction was $<10^{-6}$.
Table \ref{tab:spheres} lists the numbers of spheres which were found and placed in each volume as well as the number ratio of the two sizes.
While sorting the spheres into two classes we admitted size-deviations of $\pm20\%$ for both sizes. 
\begin{table}
\vspace{2mm}
\centering
\begin{tabular}{l|l|c}
\hline
& No of spheres & Ratio \\
Scanner & (discarded) & $10$:$20.3\,\mu$m\\
\hline
Versa 520 20x FIBRE & 45151 (173) & 8.40\\
Versa 520 4x FIBRE & 42081 (153) & 8.67\\
Versa 520 20x KIT & 64104 (274) & 8.67\\
Versa 520 4x KIT & 83291 (332) & 8.40\\
LMJ-CT FhG& 38263 (248) & 4.88\\
Click-CT 20x FhG& 7386 (39) & 7.29\\
\hline
nanotom m U WEI & 23626 (117) & 6.16\\
nanotom s HZG & 36548 (310) & 0.88\\
Easytom CCD INPG & 47101 (208) & 7.10\\
Easytom CCD RXS & 12494 (96) & 2.69\\
Easytom FP RXS & 12614 (88) & 2.75\\
Skyscan 2214 Bruker& 96851 (930) & 7.54\\
ntCT FhG& 7295 (647) & 6.79\\
\hline
\end{tabular}
\caption{Results of the ball search algorithm. Numbers of spheres which did not fit into a size population are given in brackets (cf. Fig \ref{fig:sizehisto}). The ratio only refers to the spheres which were counted.}
\label{tab:spheres}
\end{table}

The object's power spectrum $S_\mathrm{object}(u)$ is computed from $\left\vert  \mathcal{F}_{3\mathrm{D}}\{O(\vec{x})\}\right\vert^2$.
Before the Fourier transform each volume image is multiplied with an appropriate window function to suppress boundary FFT-artifacts. While comparing the peaks from harmonics of the two sphere diameters from the virtual objects' power spectrum with the measured power spectrum we discovered a certain shift in the peaks of the $20\,\mu$m spheres' harmonics. This shift was removed by slightly increasing the sphere diameter ratio, from $1:2$ to $10:20.3$, hence the assumption that the larger diameter equals $20.3\,\mu$m.

\section{Results}

\subsection{Transmission SNR spectra}

\begin{figure}[t]
\centering
\includegraphics[width=8cm]{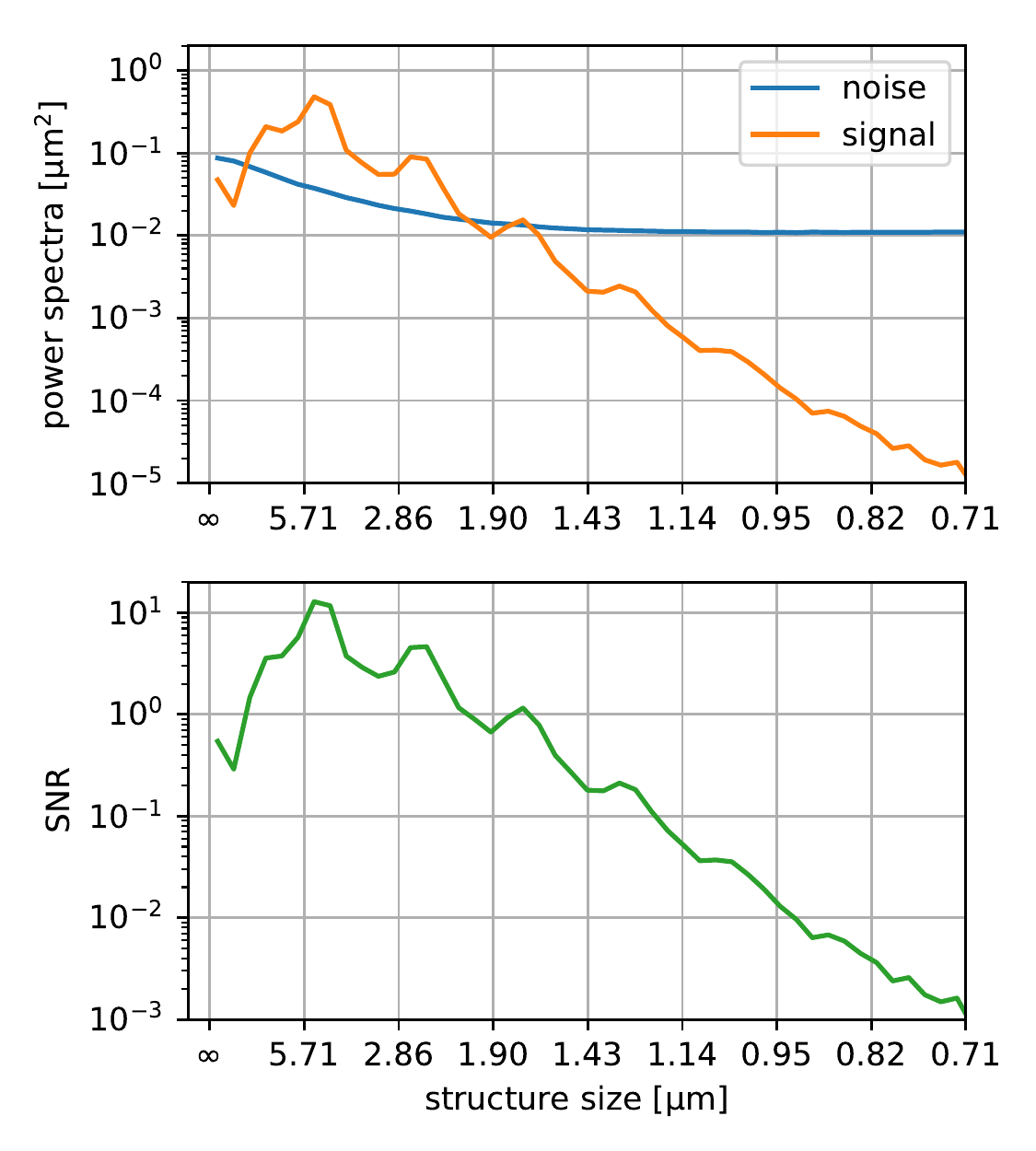} 
\caption{Noise and signal power spectra for the 20x lens of KIT's Versa 520 scanner (top) with the test phantom at the same position at which the CT scan took place. Bottom: $\mathrm{SNR_{2D}}(u)$ corresponding to $1\,$s exposure.} \label{fig:SNR-Versa-20x}
\end{figure}

Before looking at $\mathrm{SNR_{3D}}$ and $\mathrm{DE_{3D}}$, the corresponding two-dimensional measures are investigated from transmission images which are recorded for the same settings as the CT scans (see eq. \ref{eq:SNR-measure}).
Figure \ref{fig:SNR-Versa-20x} shows $S_{\mathrm{2D},\tau}(u)$, $N_{\mathrm{2D},\tau}(u)$ and $\mathrm{SNR_{2D}}(u)$ for the case of the Versa 520 (20x) scanner at KIT.
$\mathrm{SNR}(u)$ is always calculated per time (seconds for transmission images, hours for CT scans).
Like $S_{\mathrm{2D},\tau}(u)$ it shows characteristic peaks of the sphere harmonics. Based on Fig. \ref{fig:SNR-Versa-20x} it can be inferred that $\approx 1.7\,\mu$m details are detected ($\mathrm{SNR_{2D,10s}}=10$) from a $10\,$s exposure while $100\,$s exposures are required to detect $1.2\,\mu$m details.

\begin{figure*}[th]
\centering
\includegraphics[width=14.5cm]{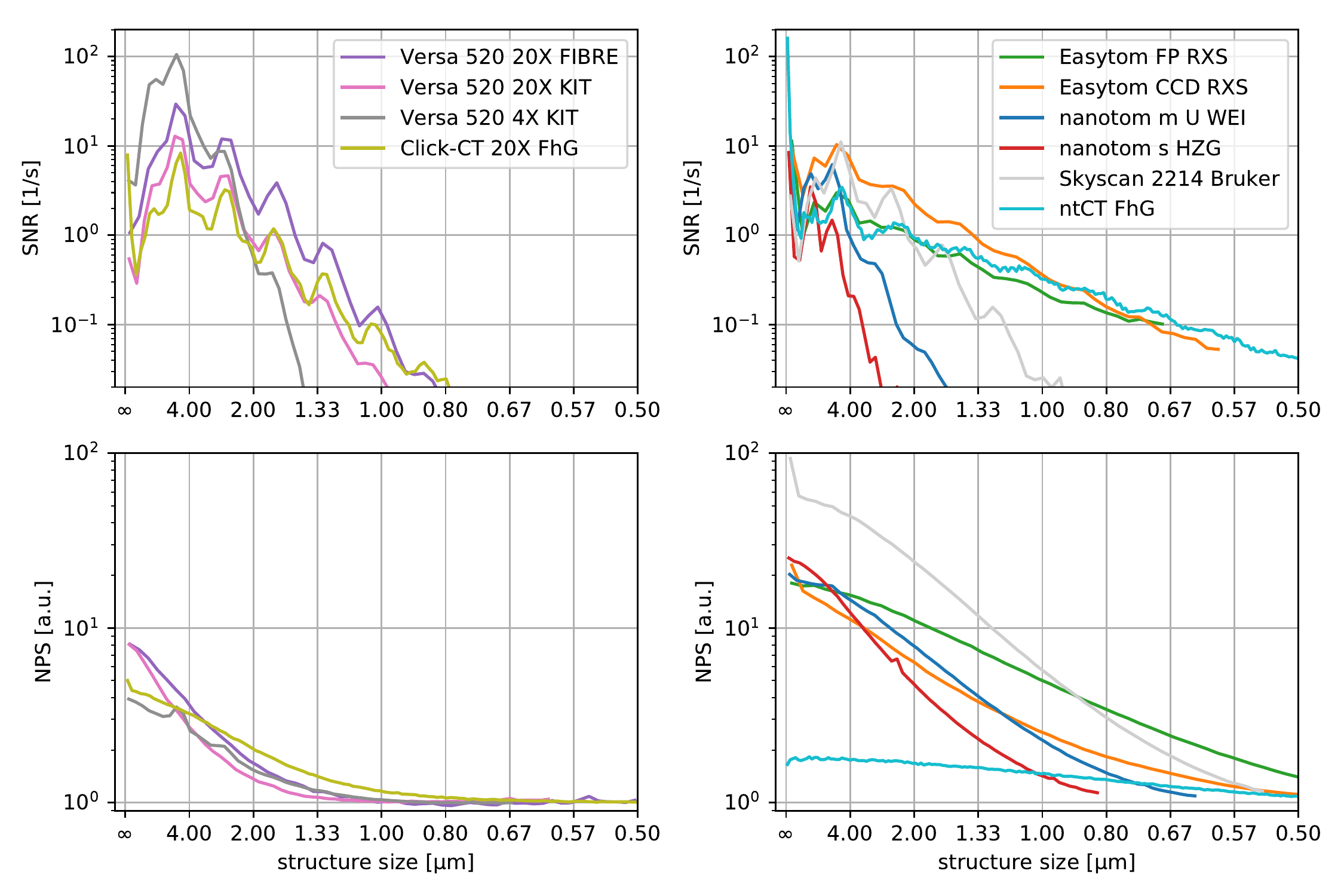} 
\caption{SNR spectra (per second exposure) from transmission 2D images for the detector- and the source-based scanners (top). Noise power spectra (bottom) are normalized with respect to the estimated NPS$(0\,\mu m)$.}\label{fig:SNR-spectra}.
\end{figure*}

Using eq. \ref{eq:conversion} we find $c\approx 7$.
For the detector-based scanners eq. \ref{eq:conversion} is valid since $\mathrm{MTF_{noise}}=0$ at $u_\mathrm{ny}$.
Yet, for many source-based scanners this condition seems not to be fulfilled (cf. Fig. \ref{fig:SNR-spectra}). Because $N_\mathrm{2D}$ does not drop to a constant value at the high frequency limit it has to be extrapolated beyond $u_\mathrm{ny}$ for estimating $c$.
The estimated X-ray conversions for all systems are given in table \ref{tab:SNR-measures}.
Note that since the values of $c$ depend on the X-ray spectrum, an increase in $c$ due to higher X-ray energies does not imply higher image quality.
\begin{table}
\centering
\vspace{2mm}
\begin{tabular}{l|l|c}
\hline
Scanner & Images & $c$ from NPS\\
\hline
Versa 520 20x FIBRE &$25\ast1\,$s & $7.0$\\
Versa 520 20x KIT &$15\ast5\,$s & $7.0$\\
Versa 520 4x KIT & $20\ast0.5\,$s & $2.9$\\
Click-CT 20x FhG & $51\ast 1\,$s & $3.4$\\
\hline
nanotom s HZG & $30\ast0.5\,$s & $24.2$\\
nanotom m U WEI & $23\ast0.75\,$s & $19.2$\\
Easytom FP RXS & $30\ast0.5\,$s & $16.2$\\
Easytom CCD RXS & $30\ast0.5\,$s & $21.9$\\
Skyscan 2214 Bruker & $30\ast 0.6\,$s & $56.0$\\
\hline
\end{tabular}
\caption{$\mathrm{SNR_{2D}}$ measurements from transmission images. Note that $N_\mathrm{2D}$ of the source-based scanners did not drop to constant values at $u_\mathrm{ny}$ and an offset value was therefore chosen manually for estimating $c$. Consequently, an error of $\pm 10\%$ should be considered for these scanners' conversions.
}\label{tab:SNR-measures}
\end{table}

Figure \ref{fig:SNR-spectra} shows $\mathrm{SNR_{2D}}$ spectra as well as $N_\mathrm{2D}$ for both types of scanners.
They should provide a first overview over the principal characteristics and performances of the scanners which will later be compared to the 3D spectra. Note, for the Easytom CCD INPG we received transmission images of a micro foam instead of the test object, yielding $c\approx 61$
(not shown in Fig. \ref{fig:SNR-spectra}).

The form factor of the microbeads (harmonics in the spectrum) is more pronounced for the detector-based systems. With the exception of the Skyscan 2214 the source-based scanners show much less harmonics in their $\mathrm{SNR_{2D}}$ spectra. This observation is explained by the close proximity of the sample to the focal spot, i.e. different magnifications of sample parts which are closer and parts which are further away from the source.

Among the detector-based scanners, $\mathrm{SNR_{2D}}(u)$ of the Versa 20x lens (KIT and FIBRE) as well as the Click-CT (which also uses a 20x lens) appear very similar. From their NPS spectra we find that both Versa 20x scans feature identical light conversions, yet the system at FIBRE has a slightly sharper detector MTF and therefore a better $\mathrm{SNR_{2D,10\,s}}=10$ at $\approx 1.5\,\mu$m while $1.7\,\mu$m can be claimed for Click-CT and Versa 20x KIT. Note that this difference mostly stems from a gain in exposure time thanks to the much shorter SDD at FIBRE. Compared to the Versa 20x both Click-CT and Versa 4x have lower light conversions. Meanwhile, the Click-CT shows a sharper detector MTF compared to all Versas. When the Versas are using their 4x lens the resulting $\mathrm{SNR_{2D}}$ is lower at high frequencies and higher at low frequencies ($\approx 2.5\,\mu$m for $\mathrm{SNR_{10s}}=10$), compared to the 20x lens.

Judging from $\mathrm{SNR_{2D}}$, three source-based systems, i.e. the RX Easytom XL, the Skyscan 2214 as well as the ntCT, clearly outperform GE's nanotom scanners. Using its CCD detector the Easytom achieves $1.4\,\mu$m detail ($\mathrm{SNR_{2D,10s}}=10$) for $10\,$s exposure while $20\,$s exposure are required when using the system's Flatpanel detector. The ntCT resolves $1.4\,\mu$m for $17\,$s exposure whereas the Skyscan 2214 requires more than one minute. Note that for $100\,$s exposure time both Easytom RXS scans (CCD and FP) could resolve $\approx 0.7\,\mu$m detail, ntCT even more. Meanwhile $\mathrm{SNR_{2D}}$ of the nanotom m scan reaches only about $4\,\mu$m detail visibility for $10\,$s exposure, the nanotom s performing even worse.

Concerning $N_\mathrm{2D}$ of the source-based scanners, the Easytom Flatpanel detector is significantly sharper compared to the CCD detector in this system (cf. Fig. \ref{fig:SNR-spectra} lower right panel). Likewise, the Flatpanel detector of the nanotom s (HZG) shows more pixel blurring than the Flatpanel in the nanotom m (U WEI). Note, the latter has twice the pixel size compared to the nanotom s, hence this difference is most likely due to the scintillator screen which shows similar X-ray conversions in both detectors. Flatpanels and CCDs in the source-based systems generally show higher X-ray conversions and less pixel blurring ($\mathrm{MTF_{noise}}$) compared to the microscope detectors. Note that NPS in the ntCT is not white noise but shows faint auto-correlation.
Most systems showed a minor focal spot drift in the $\mathrm{SNR_{2D}}$ image series which was used to compute Fig. \ref{fig:SNR-spectra} (e.g. resulting in small kinks in the NPS of the KIT 4x Versa scan).
Drift artifacts in the spectra were countered by determining SNR spectra 
from all consecutive image pairs ($K=2$ in eq. \ref{eq:SNR-measure}) and averaging over these SNR spectra.

\subsection{Volume power spectra}

\begin{figure*}[th]
\centering
\includegraphics[width=14.5cm]{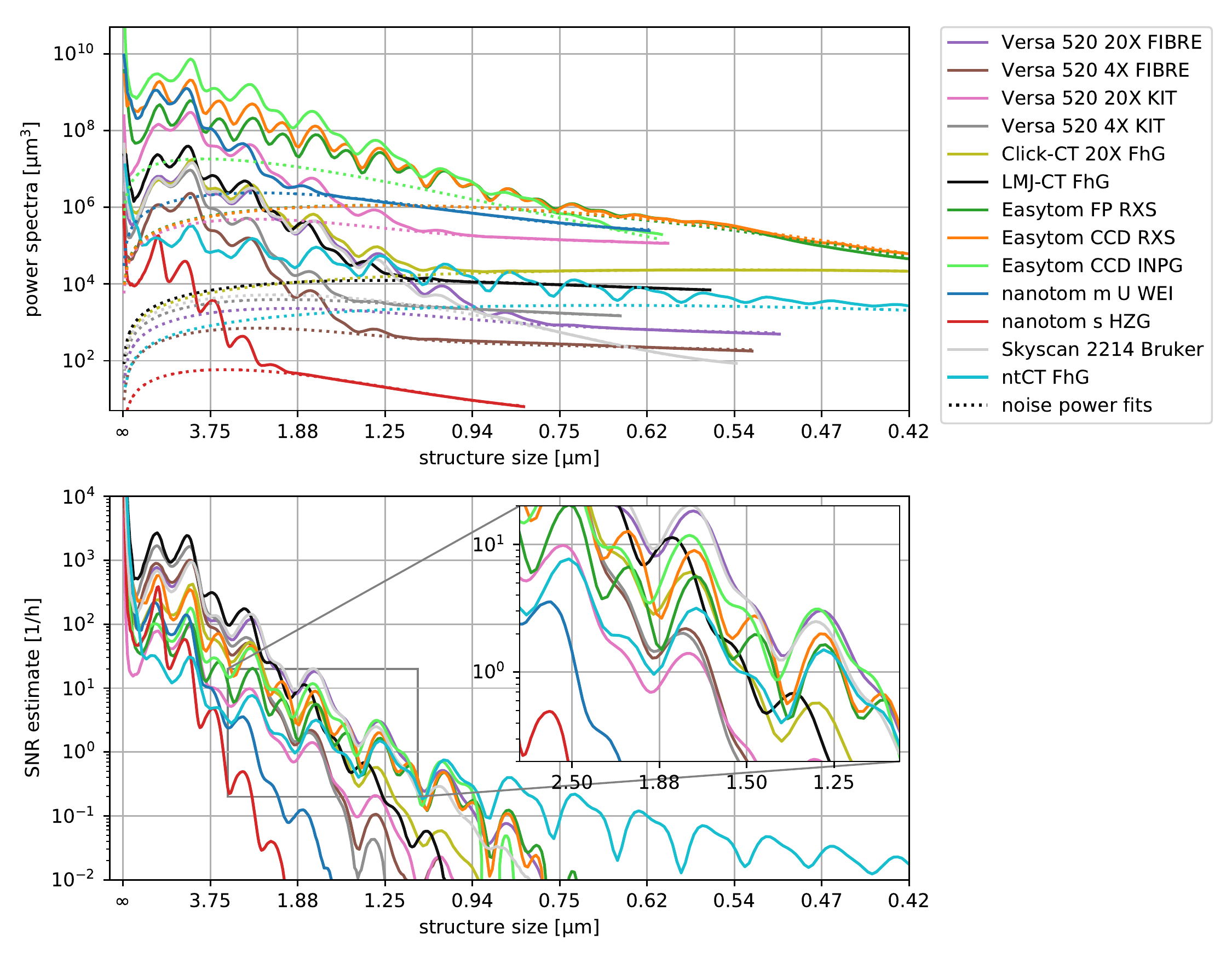} 
\caption{Radial power spectra of the measurement data from all CT scans (top). The dashed lines display estimates of the noise power spectra according to eq. \ref{eq:NPS} (model). Note that the power spectra are not normalized, their scaling is arbitrary. Estimated volume $\mathrm{SNR_{3D}}(u)$ spectra per hour of measurement time (bottom).}\label{fig:ct-Power-spectra}
\end{figure*}

Figure \ref{fig:ct-Power-spectra} shows the data power spectra $D_\mathrm{3D}$ which were computed from CT images of our test phantom by the routine described above. The model fit of $N_\mathrm{3D}$ is shown along with data power spectra.
It is computed according to eq. \ref{eq:NPS} using the values of $c$ from table \ref{tab:SNR-measures} as constants. The fit is very robust for those scans which display pure noise in a high frequency band. Note that for the Easytom RXS scans this band is somewhat short, therefore the fit is less certain. Particularly, the Easytom INPG scan did not have a pure noise band in its spectrum at all. Therefore, corresponding $\mathrm{SNR_{3D}}$ has to be discussed with a lot of care. Signal and noise spectra appear distributed over a broad range of magnitudes ($10^1-10^{10}$).
This is mainly due to differences in phase contrast and attenuation strengths.

$\mathrm{SNR_{3D}}(u)$ spectra are computed according to eq. \ref{eq:3DSNR} (cf. Fig. \ref{fig:ct-Power-spectra} bottom panel). Their scaling is absolute and normalized with respect to the total scan time (cumulative time of actual exposures).
Consequently, $\mathrm{SNR_{3D}}$ emphasizes the best scan quality per hours of exposure time.
Because of the strong modulation due to the object spectrum, interpreting resolving power from SNR is difficult. 
How much scan time requires each scanner for resolving $1.5\,\mu$m details, i.e. $\mathrm{SNR_{3D,1.5\,\mu m}}=10$?
The Skyscan 2214 and Versa 20x FIBRE scans require $3.1\,$hrs whereas the Easytom scans take slightly longer: $3.3\,$hrs, $4.3\,$hrs and $6.7\,$hrs for the Easytom CCD INPG, the Easytom CCD RXS and the FP RXS respectively. With $10\,$hrs scan time Fraunhofer Click-CT, ntCT and LMJ-CT are all able to resolve this level of detail. Within this time frame ($10\,$hrs) the two Versa 4x scans as well as the Versa 20x KIT scan resolve slightly less detail ($\approx1.6\,\mu$m) whereas only $\approx 3.6\,\mu$m and $\approx 2.5\,\mu$m are detected by the nanotom s (HZG) and nanotom m (U WEI) respectively.
Note that for other structure-detail sizes, the systems' performances differ.

\subsection{Volume modulation transfer functions}

\begin{figure*}[th]
\centering
\includegraphics[width=14.5cm]{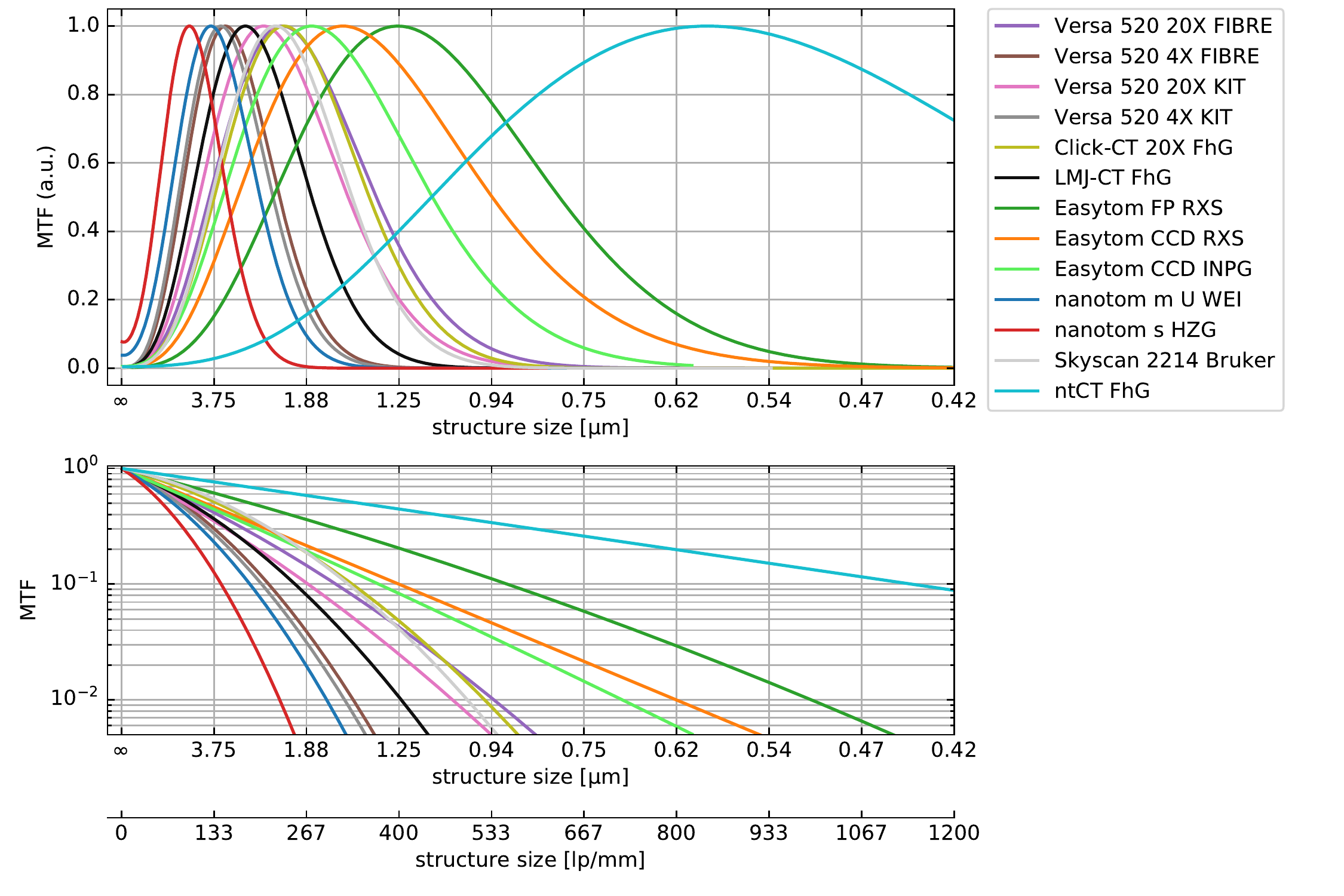} 
\caption{Model fits of $\mathrm{MTF_{3D}}$ for all scans (top). Due to normalizing the peak amplitudes to unity, the zero fequency now differs according to the ratio between attenuation and phase strength. Bottom: $\mathrm{MTF_{3D,blur}}(u)$.}\label{fig:model-fit-ct}
\end{figure*}

Model fits of $\mathrm{MTF_{3D}}$ including blur and phase contrast are displayed in Fig. \ref{fig:model-fit-ct} for all scans along with the resolving power $\mathrm{MTF_{3D,blur}}$.
Because phase contrast --and not attenuation-- is the dominant X-ray interaction, all $\mathrm{MTF_{3D}}$ are peaked. Their maxima are set to unity for the sake of simplicity.
$\mathrm{MTF_{3D,blur}}(u)$ includes blurring by source and detector.
Here, we shall examine MTF at its $10\,\%$ threshold which is a common measure for evaluating spatial resolution.
For the source-based systems the latter is mostly influenced by the focal spot width.
Unlike $\mathrm{SNR_{3D}}$ spectra, $\mathrm{MTF_{3D}}$ can be strictly interpreted with regard to spatial resolution, because it is independent from noise, object shape and exposure time.

For the two nanotom scans the phase contrast is cut off by strong image blur at relatively low spatial frequencies, hence their imaging capability is limited to structures well above $1\,\mu$m size.
Note that phase contrast enhances the imaging capability of a system beyond the mere resolving power in terms of $\mathrm{MTF_{3D,blur}}$.
The latter is lowest for the nanotom scans with the nanotom m resolving slightly smaller structures compared to the nanotom s ($2.9\,\mu$m and $3.7\,\mu$m respectively).
On the other hand, the three Easytom scans show the best $\mathrm{MTF_{3D}}$ among the commercial scanners.
The Easytom RXS scan with the Flatpanel detector has the best resolution, covering even structures below $1\,\mu$m size ($\mathrm{MTF_{3D,blur,10\%}}\approx 0.9\,\mu$m) while the scans with the CCD detector reach $\mathrm{MTF_{3D,blur,10\%}}\approx 1.25\,\mu$m. Because of its lower energy and shorter propagation length, $\mathrm{MTF_{3D}}$ of the INPG Easytom CCD scan shows a peak which is shifted towards lower spatial frequencies compared to the peaks of the RXS scans (higher energy and longer propagation). Despite these differences, all three scans feature high spatial resolution ($\mathrm{MTF_{3D,blur,10\%}}$).

Among the detector-based systems the Versa 20x scan from FIBRE and Click-CT show the best $\mathrm{MTF_{3D, blur}}$. Meanwhile the phase contrast peak of the 20x scan from KIT is shifted towards larger structural sizes despite using a longer propagation and the same energy as the 20x scan from FIBRE. This difference in $\mathrm{MTF_{3D}}$ between the 20x scans at KIT and FIBRE is likely due to their different resolving powers: The Versa at FIBRE reaches $\mathrm{MTF_{3D,blur,10\%}}\approx 1.6\,\mu$m while only $\approx 1.9\,\mu$m are obtained at KIT. Among the detector-based systems Click-CT has the best $\mathrm{MTF_{3D,blur}}\approx 1.5\,\mu$m, whereas the LMJ-CT resolves only $\approx 2.0\,\mu$m. 
The transfer function of the source-based Skyscan 2214 resembles the MTF of the Versa 20x at KIT while its resolving power is visibly better ($\approx 1.5\,\mu$m). 
$\mathrm{MTF_{3D}}$ of the two Versa 4x scans cover a somewhat lower frequency band ($\approx 2.7\,\mu$m).
$\mathrm{MTF_{3D}}$ and $\mathrm{MTF_{3D, blur}}$ of the ntCT extend beyond this figure. The phase contrast peak of the MTF is centered around $0.6\,\mu$m structure size. Judging from $\mathrm{MTF_{3D,blur,10\%}}$ the ntCT resolves even structures down to $0.4\,\mu$m size.

\subsection{Detection effectiveness}

\begin{figure*}[th]
\centering
\includegraphics[width=15.5cm]{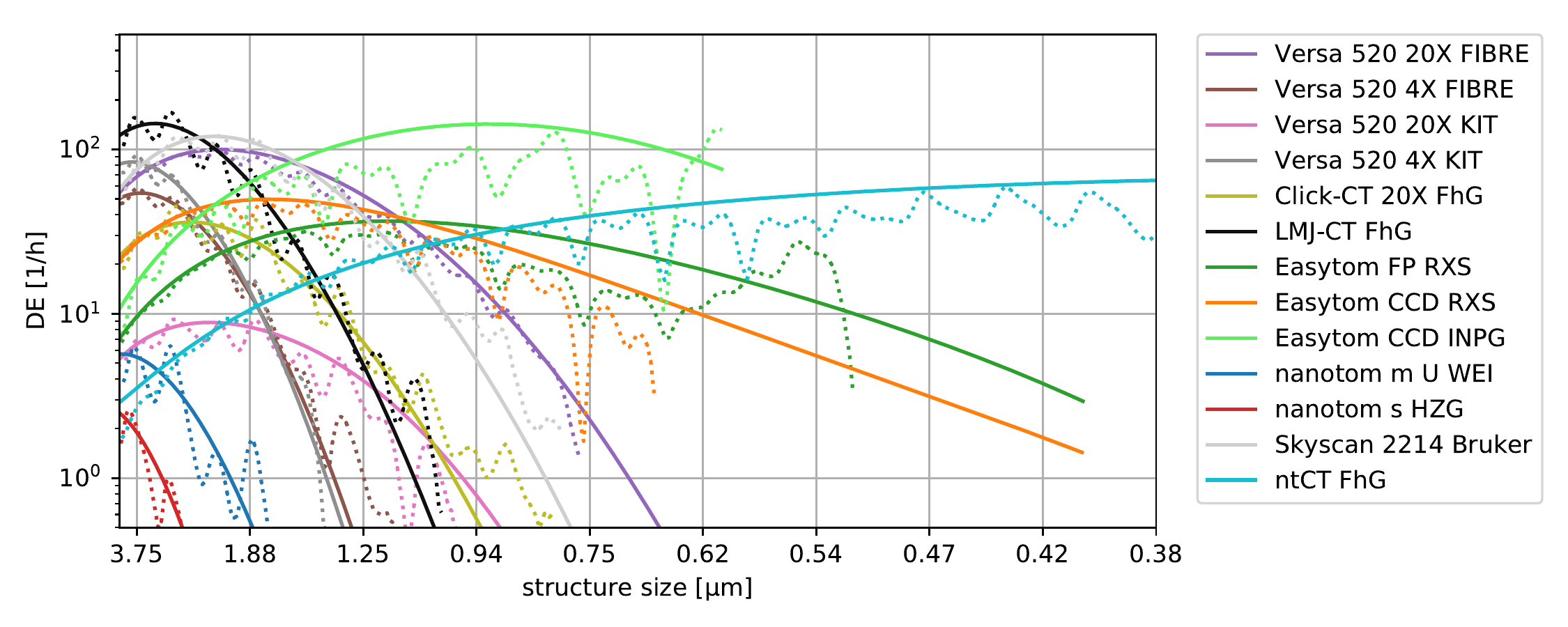} 
\caption{Data (dotted line) and model (solid line) for Detection Effectiveness in all scans.}\label{fig:snr-factor-all}
\end{figure*}

Finally, $\mathrm{DE_{3D}}$ is computed by normalizing the estimated $\mathrm{SNR_{3D}}(u)$ (Fig. \ref{fig:ct-Power-spectra}) with respect to $S_\mathrm{object}(u)$. The resulting detection effectiveness is shown as data and model fit in Fig. \ref{fig:snr-factor-all}. By "model" we mean the fit for $b^2\cdot|\mathrm{MTF_{3D}}(u)|^2$ (Fig. \ref{fig:model-fit-ct}) divided by the $N_\mathrm{3D}$ model fit (cf. Fig. \ref{fig:ct-Power-spectra} top panel).
Similar to $\mathrm{MTF_{3D}}(u)$, $\mathrm{DE_{3D}}$ is peak-shaped. Here, unlike in Fig. \ref{fig:model-fit-ct}, $\mathrm{MTF_{3D}}$ was not normalized with respect to its peak height. Therefore, $\mathrm{DE_{3D}}$ contains the actual phase-strength of each scan (in addition to blur and noisiness). In total, $\mathrm{DE_{3D}}$ indicates how well structures of a certain size can be detected by a certain system and setting during a defined scan time. 

While MTF favored the Easytom scans with the better resolving power and stronger phase contrast (Fig. \ref{fig:model-fit-ct}), DE is much higher for the Easytom INPG which features outstanding short $z_\mathrm{SOD}$ and scan time.
The detection effectiveness of the two longer RXS Easytom scans clearly does not favor one over the other. Only for structures smaller than $1.1\,\mu$m the Flatpanel yields a slightly better DE compared to the CCD detector. Note, the $N_\mathrm{3D}$ model fit leads to slightly over-estimate DE for the high frequencies, hence both RXS settings might actually be equal.
For structures smaller than $0.6\,\mu$m the ntCT has the highest DE, extending far into the sub-micrometer range.

Among the detector-based systems the LMJ-CT scan reaches the highest DE, but only for structures larger than $2.4\,\mu$m. For smaller structures the Versa 20x scan at FIBRE as well as the Bruker Skyscan 2214 yield a better DE. The two Versa 20x scans again show significantly different DE. The FIBRE scan is an order of magnitude better than the KIT scan. This effect results mostly from the much shorter SDD but also from the higher resolving power of the FIBRE scan (cf. Fig. \ref{fig:model-fit-ct}). Meanwhile the 4x scans from the two Versa scanners yield almost identical DE, whereas Click-CT yields a performance in between the two Versa 20x scans.

The two nanotom scans clearly show the lowest DE. Hereby, the nanotom m achieves better results than the nanotom s. The DE peak of the nanotom scans is two orders of magnitude weaker than the DE of the LMJ-CT, and one order of magnitude weaker than the Click-CT.

\section{Discussion}

\subsection{Transmission power spectra}

A lot can be deduced about the capabilities and the performance of sub$\mu$ CT scanners from transmission images alone. From $N_\mathrm{2D}$ and and $\mathrm{SNR_{2D}}$ spectra we could evaluate and compare the resolving power, the signal strength and the noisiness of these machines simply from a series of sample images. In order to evaluate the signal (not noise) MTF and to estimate the latter independently for detector and source, additional images of a slanted edge can be used (see Appendix).
Already at this level (projection) we find significant differences between source-based and detector-based scanners.

While the noise spectra of detector-based scanners look relatively similar, featuring a steep MTF, $N_\mathrm{2D}$ of the source-based systems show higher X-ray conversions with different resolving powers. Scintillators and screen thicknesses as well as the choice between Flatpanel and CCD all contribute to those detectors' performances.

\subsection{Volume power spectra}

As it would be expected, volume $\mathrm{SNR_{3D}}$ spectra of the 13 scans mainly reproduced the results of the transmission $\mathrm{SNR_{2D}}$ spectra.
One flaw of this study shows in the number ratios of $10\,\mu$m spheres vs. $20.3\,\mu$m spheres. While some scans come very close to the theoretical value, others show much lower values. The most extreme example is the test sample from the nanotom s scan which contained more large than small spheres. Also the two RXS scans show a significantly lower portion of small spheres (most likely the mixing of the two powders was incomplete). While these deviations certainly alter the object spectrum and thereby SNR,
MTF and DE should remain unaffected by this cause.
Even though the envelope of the object spectra from the nanotom m and s does not appear different, more care should be taken to prevent demixing.

SNR spectra (transmission and volume) scale with the object's power spectrum.
For our test phantom, which was designed to characterize all sub$\mu$ CT scanners, we compared $\mathrm{SNR_{3D}}$ of all scans at $\approx 1.5\,\mu$m structure size, finding that two scans performed best: Both Versa 520 20x FIBRE and Skyscan 2214 could resolve this level of detail in $3.1\,$hrs scan time.
Yet, this result is only a tiny part of the full imaging capabilities which can be estimated from $\mathrm{SNR_{3D}}$ and $\mathrm{DE_{3D}}$.

In retrospect, using somewhat coarser spheres and estimating the potential imaging quality of the systems in terms of $\mathrm{SNR_{3D}}$ was a good choice, because a. the two types of quasi mono-disperse spheres well allowed for extrapolating $\mathrm{SNR_{3D}}$, $\mathrm{MTF_{3D}}$ and $\mathrm{DE_{3D}}$ into the sub-micrometer range, and b. smaller spheres would have excluded the nanotom systems entirely from this study.

On the one hand, the volume transfer function $\mathrm{MTF_{3D}}$ has been shown to reflect the systems capacities in terms of spatial resolution and hence phase contrast. Note, MTF does not depend on the object structure, nor does it include noisiness: Any MTF is shape- and time-independent.

On the other hand, $\mathrm{SNR_{3D}}$ spectra reflect the resolving power as well as the noisiness of a scan, yet they inevitably contain the object structure and are therefore shape-dependent by definition. The newly defined detection effectiveness (SNR normalized with respect to the object spectrum) is the only measure which includes modulation transfer and noisiness (SNR) while being independent of the object structure. Generally, $\mathrm{DE_{3D}}$ is very weak for the nanotom scanners, more or less strong for the detector-based systems (depending on the resolving power of their detector) and very strong for the Easytom XL and ntCT. 

Note that SNR and DE are invariant under any Fourier-filter, such as Paganin-type phase retrieval. Therefore even though the raw CT data does not show linear attenuation coefficients, resulting SNR and DE spectra can be interpreted as if they did.

\subsection{Operator-choices}

Most operators were aware of phase contrast being the main signal source in X-ray sub$\mu$ imaging, particularly for weakly attenuating objects. All scans feature different propagation lengths: While LMJ-CT, Versa 20x KIT and the Skyscan 2214 scan by Bruker feature relatively large $z$, a shorter propagation is used for Versa 20x FIBRE, Click-CT and Easytom INPG.
The latter feature better DE, whereas scans with a larger SDD have a better MTF.
This difference is explained by the phase contrast strength increasing linearly, while intensity decreases quadratically with $z_\mathrm{SDD}$. Note that X-ray phase contrast further scales with the X-ray energy $\varpropto E^{-2}$.
In some cases $z_\mathrm{SOD}$ was limited, e.g. the test object used by RXS had a glue bulb sealing the glass capillary which was $2-3\,$mm large, thus preventing the operator from shortening SOD below $3\,$mm (in contrast to the $1.8\,$mm for the Easytom INGP).
The Versa 4x scans make a special case: Both at KIT and at FIBRE the object magnification was $M>2$, yet remained moderate (5.2 and 6.5 respectively) thus allowing for relatively long SOD. Since the focal spot in this setting is similar but slightly smaller than the detector resolution (see Appendix) this particular object position allows for undercutting both contributions yielding an MTF with strong phase contrast and good spatial resolution at relatively short scan times.

The strong DE of the LMJ-CT mainly stems from the superior brilliance of the X-ray source. Note that in addition to the high target power, the Gallium anode in this system emits quasi monochromatic X-rays which always yield higher DE compared to polychromatic imaging. Meanwhile the high DE of the Easytom INPG scan, in particular with respect to the lower DE of the two RXS scans (which feature better MTF), stems from a number of interesting operator choices.
Setting the voltage to $40\,$kV instead of $100\,$kV has two consequences: a. Low energies contribute significantly more to the signal, whereas high energies contribute more Poisson noise than signal thereby worsening SNR and DE, and b. Using $100\,$kV acceleration produces a smaller focal spot, i.e. better resolution, by narrowing the electron interaction bulb in the transmission target ($1\,\mu$m tungsten).

\subsection{Technicalities of the detector-based scanners}

Compared to the 4x lens, the 20x lens increases spatial resolution which is limited by noise and by the X-ray conversion $c$. In that respect, the 20x lens has a double advantage over the 4x lens: Its lens has a higher resolving power and a higher light collection efficiency (the resulting conversion $c$ is about twice the value of the 4x lens including likely different scintillator thicknesses). It is surprising that both operators - when given the choice - first selected the 4x lens for imaging the test phantom because of its larger field-of-view. Using the 4x lens also implies shorter exposures due to coarser pixel sampling and higher stopping power of the thicker scintillator screen. Note, the latter mostly adds high-energy X-rays to the detection and does not necessarily benefit image quality (SNR).

The Versa 20x FIBRE scan shows a better $\mathrm{MTF_{3D}}$ than the corresponding scan from KIT, even though the latter should have less penumbral blurring due to a stronger source demagnification. $\mathrm{MTF_{2D}}$ indicates that the 20x lens at FIBRE resolves more detail (yet, we determined $1.0\,\mu$m for the 20x at KIT). The two systems show the same X-ray conversion. Additionally, the FIBRE source may have a smaller focal spot, but this was not measured in this study. The LMJ-CT scan suffered from focal spot drift which was corrected through linear drift correction. The latter resulted in a slight deformation of the microbeads. Not visible to the eye, this deformation became visible in the signal spectrum and was corrected by a slight elongation of the spheres in the virtual object image (see section~\ref{sec:Volume}).

The Click-CT and the Versa 20x detectors do not employ the exact same technology. The Versa system has a light conversion which is twice as high, yet it resolves less details than the Click-CT (MTF). While the Click-CT uses a thin transparent scintillator crystal as it is common at synchrotron beamlines \cite{Koch:1998}, the Versa detector presumably employs a poly-crystalline CsI screen, which has a higher light yield, yet less stopping power. The Versa detector most likely uses a 20x lens with lower numerical aperture (NA) compared to Click-CT (NA 0.75), a decision which makes active lens-focusing, as it is done in the Click-CT, obsolete, yet reduces the overall light conversion. In sum, these differences seem to compensate each other, while shortening SDD remains decisive.

\subsection{Technicalities of the source-based scanners}

These systems show much different performance in terms of SNR, MTF and DE. These differences mostly stem from different focal spots sizes. The latter are determined by the corresponding X-ray sources whose spot size can be altered, e.g. by choosing different cathodes, different acceleration voltages and transmission anodes of different thickness. From our obersvations we roughly estimate a focal spot of $3-4\,\mu$m size for the nanotom sources whereas (according to the data sheet) the Easytom source with LaB6 cathode has a focal spot of $\approx 0.6\,\mu$m which would be a factor 6 smaller. The ntCT employs the Nanotube source by the Swedish company Excillum which reaches a focal spot below $0.3\,\mu$m (knife edge scan) \cite{Nachtrab:2014, Fella:2018}. In terms of $\mathrm{MTF_{3D,blur}}$ Easytom and ntCT achieve $0.9\,\mu$m and $0.4\,\mu$m respectively. Note that while resolution and MTF relate to focal spot width, those two measures are not the same. While the ntCT employs a photon counting detector (Dectris Säntis CdTe), the Easytom offers two options: a tapered CCD or a Flatpanel detector.
The latter has a better resolution, while sharing similar $\mathrm{SNR_{3D}}$ and $\mathrm{DE_{3D}}$ spectra with the CCD.
The Dectris Säntis detector is not yet released and therefore employs an experimental calibration, i.e. for the low-energy threshold ($6\,$keV).
 Note that the Easytom INPG scan employed a different CCD detector than the Easytom CCD RXS scan. The former used a QuadRO by Princeton Instruments with $100-150\,\mu$m CsI scintillator (now discontinued) whereas the newer model might employ less CsI due to its smaller pixels.
The reason why the Easytom CCD performs almost as well as the Flatpanel (Varian) which certainly has a thicker scintillator remains uncertain. Probably, the acceptance in terms of X-ray energies in both detectors is not the same, hence the CCD detector likely has a more favorable energy window compared to the Flatpanel even if their total light conversions are similar. 

At first glance, the Skyscan 2214 is similar to the RX Easytom: they both employ a Hamamatsu transmission tube and tapered CCD detectors. Yet, their SNR, MTF and DE are different. The Bruker scan emphasizes phase contrast at the expense of focal spot size and the result matches the quality of the Versa 20x FIBRE. The Easytom scans achieve higher resolution by employing LaB6 cathodes in their X-ray tube, whereas Bruker used a tungsten filament (note that LaB6 is also available for the Bruker system). The Easytom uses Cesium-Iodide (CsI) as scintillator material (FP and CCD), while the Ximea CCD detector in the Skyscan 2214 uses Gadolinium-Oxysulfide (GdOS). If and how exactly these choices influence the outcome
requires further examination. The Skyscan and Easytom scans might be considered a set of performances, while one particular scan is realized by a set of options. 
While the best detector-based scan (Versa 520 20x FIBRE) seems to show the performance limit of this technology, at least three source-based systems perform beyond this limit. Hereby, the Skyscan 2214 likely obtained the best SNR at the cost of a slightly worse resolution.

\section{Conclusion}

This report is meant to bring quantitative arguments into a discussion which has recently been very lively but also very qualitative. It began with the assumption that all commercial sub$\mu$ CT scanners shared similar performances in terms of resolving power and image quality and that experienced operators would always choose the optimal settings.
This study shows that for each system two kinds of optima can be achieved: a. The optimal resolving power (best MTF), or b. the optimal signal-to-noise ratio for a given object and scan time (best SNR). While our results reveal technical limits for source- and for detector-based scanners, they show that operator-choices were either made to achieve best MTF or best SNR.
Concerning MTF, source-based scanners have the upper hand over detector-based systems. Yet, only if a. the operator chooses the optimal settings (most importantly anode voltage and source-detector distance), and b. the X-ray focal spot is smaller than the optical resolution limit of the microscope lens in state-of-the-art detector-based systems. We estimated the latter to be $\approx 1\,\mu$m, therefore only smaller focal spots achieve a better MTF. In that respect the ntCT ($0.4\,\mu$m) certainly qualifies as the gold standard for sub$\mu$ X-ray imaging.

Concerning SNR, this study showed that setting up a high-resolution CT scan with optimal parameters is a highly complex task. Consequently, the need for experimental optimization and guided scan parameterization is very high. Setting the anode voltage too high may decrease SNR (despite improving resolution), while choosing a less magnifying lens in a detector-based system may reduce light conversion as well as spatial resolution. Note that SNR spectra, most importantly those based on simple transmission images, also include the resolving power of a system. Meanwhile MTF defines resolution as an object- and time-independent frequency cutoff, i.e. at $\mathrm{MTF}_{10\%}$. In most real-life cases detection is not limited by MTF but by SNR which includes the object's power spectrum as well as the noisiness of the scan. Therefore, the latter makes a perfect tool for carrying out an automated optimization for any given setup and task. Optimal $\mathrm{SNR_{2D}}$ implies optimal $\mathrm{SNR_{3D}}$ which in turn implies optimal $\mathrm{DE_{3D}}$. Note, here volumetric descriptors were estimated from single volume images. Consequently, $\mathrm{SNR_{3D}}$ and $\mathrm{DE_{3D}}$ both relied on model fits of the CT noise power spectrum which ideally would have been measured.
$\mathrm{SNR_{3D}}$ and $\mathrm{DE_{3D}}$ serve for defining and comparing any scanners' performance. Measuring $N_\mathrm{3D}$ for CT scans could work the same way as measuring $N_\mathrm{2D}$ from transmission images.
Therefore, a special CT scan acquisition has to be employed which unfortunately was not available for the commercial scanners.
MTF certainly provides a basis for quantitative comparison of imaging systems' performances, from a point of view which is independent of object and exposure time. Yet, it is precisely because of these properties that MTF should not be employed solely for system optimization and comparison.
Unlike $\mathrm{MTF_{3D}}$, $\mathrm{SNR_{3D}}$ and $\mathrm{DE_{3D}}$ simultaneously include all effects influencing image quality (noisiness, blur and phase contrast). Given a reproducible test object, their measure is an absolute, quantitative scale. For system comparison and optimization they should be measured as well as whenever new systems are introduced.

\section*{Acknowledgments}

We are much grateful to Jochen Joos from Karlsruhe Insititute of Technology, to Solene Valton from Rayons X Solutions , to Antonia Ohlmann and Dominik Müller from University Würzburg and to Franziska Vogt from Bauhaus-Universität Weimar who dedicated their time and their instruments for obtaining the data for this report and for discussing the results. Theobald Fuchs and Alison Haydock shall be thanked for discussion and for proof-reading the manuscript. Benedikt Sochor was a great help in preparing the samples. Miriam Resch and Tilman Schwemmer helped testing the sphere search algorithm.
This research did not receive any specific grant from funding agencies in the public, commercial, or not-for-profit sectors.

\section*{Author Contributions}

S.Z. and M.U. equally contributed the bulk of this study.
M.U. took the lead in developing and testing the physical models, and wrote the code for data processing, analysis and visualization. M.U. is the author of pyXIT.
S.Z. took the oversight and the leadership for the research, as well as preparation of the draft and writing of the manuscript. S.Z. also coordinated and supervised the experiments.
S.Z. and M.U. take equal responsibility for the interpretation of the results.
R.H. and S.Z. formulated the research idea and chose the participants.

S.Z. and R.S. designed the test object.
O.F., C.F., B.Z.-P., P.L. and W.DB. performed the experiments and provided input and commentary as well as critical review of the draft.

\bibliographystyle{elsarticle-num}

\appendix
\section{Signal artifact correction}
Computation of $\mathrm{SNR_{2D}}$ involves a correction term $A(u)$ corresponds to artificial signal contributions (e.g., arising from the detector dark current or brightfield noise power spectra, see eq. \ref{eq:SNR-measure}).
$A(u)$ is computed from two image series: a. $K_\mathrm{flat}$ flatfield images $I_\mathrm{flat,j}$, $j\in 1\dots K_\mathrm{flat}$; b. $K_\mathrm{dark}$ dark-current images $I_\mathrm{dark,j}$, $j\in 1\dots K_\mathrm{dark}$.
In order to evaluate the resulting artificial signal, both series are converted to transmission image series (see eq. \ref{eq:normalize}):
\begin{equation}
d_\mathrm{flat,j}= -\ln\left[\frac{I_\mathrm{\tau,avg}-I_\mathrm{dark,avg}}{I_\mathrm{flat,j}-I_\mathrm{dark,avg}}\right]\label{eq:flat-series}
\end{equation}
\begin{equation}
d_\mathrm{dark,j}= -\ln\left[\frac{I_\mathrm{\tau,avg}-I_\mathrm{dark,j}}{I_\mathrm{flat,avg}-I_\mathrm{dark,avg}}\right]\label{eq:dark-series}
\end{equation}
Here, $I_\mathrm{\tau,avg}$ is the average measured intensity image. From eq. \ref{eq:flat-series} and \ref{eq:dark-series} individual ($D_{\mathrm{flat},\,j}$) and average power spectra ($D_\mathrm{flat,\,avg}$) are computed, yielding two correction terms
\begin{equation}
A_\mathrm{flat}(u)=\frac{\left\langle D_{\mathrm{flat},\,j}\right\rangle-D_\mathrm{flat,\,avg}}{(1-K^{-1}_\mathrm{flat})K_\mathrm{flat}}
\end{equation}
\begin{equation}
A_\mathrm{dark}(u)=\frac{\left\langle D_{\mathrm{dark},\,j}\right\rangle-D_\mathrm{dark,\,avg}}{(1-K^{-1}_\mathrm{dark})K_\mathrm{dark}}
\end{equation}
Subtraction of $A(u)=A_\mathrm{flat}(u)+A_\mathrm{dark}(u)$ yields an artifact free SNR (cf. eq. \ref{eq:SNR-measure}).

\section{MTF from a slanted edge}
For the Zeiss Versa 4x and 20x settings at KIT two series of six transmission images from a slanted tungsten knife edge (thickness $25\,\mu$m, $\approx 7^\circ$ tilt to ensure oversampling) were recorded at increasing magnification, i.e. at six different positions between source and detector, while keeping the source-detector distance $z_\mathrm{SDD}$ constant.
Eq. \ref{eq:MTF-model} was used to create a model fit to each MTF. 
At each position we obtained an accurate measure of the pixel sampling by shifting the slanted edge sideways, i.e. a precise horizontal displacement ($0.2\,$mm or $0.5\,$mm) and counting the latter in pixels. 
By assuming a known detector pixel size we obtained $z_\mathrm{ODD}$ as well as the $z_\mathrm{SOD}$ at every sample position.
For the 20x and the 4x lens source- and detector blurring were thus estimated in a consistent manner.
\begin{figure}[hbtp]
\centering
\includegraphics[width=8.0cm]{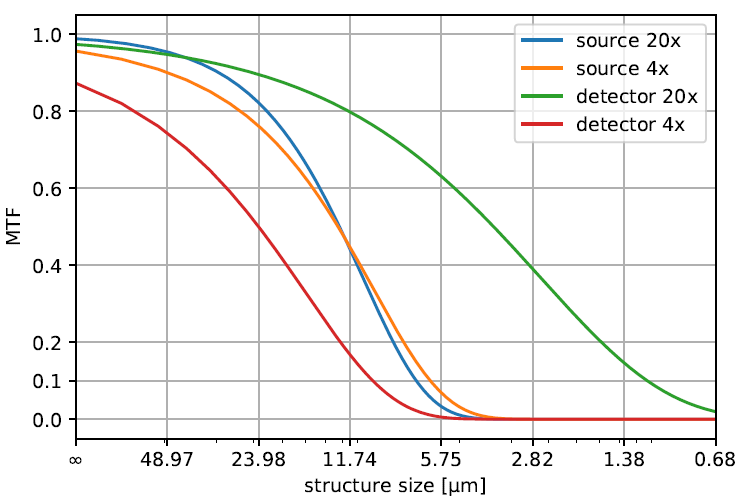}
\caption{Model fit for source- and detector MTF for the KIT Zeiss Versa 4x and 20x.}\label{fig:source-det-blur-KIT}
\end{figure}
Figure \ref{fig:source-det-blur-KIT} displays these two contributions as they were obtained from the fitting. Visibly the $\mathrm{MTF_{source}}$ curves are similar but not identical for the two fits. We believe the curve from the 4x lens to be more accurate because the edge images by the 20x setting contained additional phase contrast which artificially raised the sharpness of the edge. Concerning the two lenses, the 20x reaches $1\,\mu$m resolution ($\mu\approx 0.625\,\mathrm{px}$, $\sigma\approx 0$) while the 4x lens only yields $9-10\,\mu$m resolution ($\mu\approx 0.588\,\mathrm{px}$, $\sigma\approx 0.667\,\mathrm{px}$) at $10\%\mathrm{MTF}$. The focal spot of the X-ray source is $6-7\,\mu$m. These results were compared to the volume $\mathrm{MTF_{blur}}(u)$ at the scan position of the CT scan on the same system (KIT), producing a very good match.

\end{document}